# Low-temperature Fabrication of Highly-Efficient, Optically-Transparent (FTO-free) Graphene Cathode for Co-Mediated Dye-Sensitized Solar Cells with Acetonitrile-free Electrolyte Solution

Ladislav Kavan[1,2*], Paul Liska[1], Shaik M. Zakeeruddin[1] and Michael Graetzel[1]

[1]*Laboratory of Photonics and Interfaces, Institute of Chemical Sciences and Engineering, Swiss Federal Institute of Technology, CH-1015 Lausanne, Switzerland*

[2]*J. Heyrovský Institute of Physical Chemistry, v.v.i., Academy of Sciences of the Czech Republic, Dolejškova 3, CZ-18223 Prague 8, Czech Republic*

*e-mail: kavan@jh-inst.cas.cz

**ABSTRACT**

Propionitrile electrolyte solutions mixed with sulfolane or with 1-ethyl 3-methyl imidazolium tetracyanoborate (ionic liquid) are optimized for $Co(bpy)^{3+}/Co(bpy)^{2+}$-mediated DSCs working at low illumination intensity. Highly-active cathode catalysts based on graphene oxide, either pure or mixed with graphene nanoplatelets or with stacked graphene fibers, can be prepared at temperatures $\leq 200^oC$. The catalytic layers are well adhering to the substrates, i.e. to FTO or to stainless-steel surfaces, both the flat steel sheet and the steel wires in woven fabric consisting of transparent polyester (PEN) fibers in warp and stainless steel wires in weft (Sefar B23). The dye-sensitized solar cells with various cathodes, fabricated either from Pt or from optimized graphene-based catalysts, and supported by either FTO or by stainless-steel/PEN fabric show similar solar conversion efficiencies between 6.9 and 7.9 % at 0.25 sun illumination.

**KEYWORDS**: dye sensitized solar cell; electrochemical impedance spectroscopy; stainless-steel; woven fabric; Co-mediator

## 1. Introduction

The dye sensitized solar cell (DSC) also called the Graetzel cell [1,2] is an efficient, low-cost photovoltaic device achieving competitive parameters on the lab-scale, but its pervasive commercialization still requires some improvements. A classical DSC employs a sensitized $TiO_2$ photoanode supported by an F-doped $SnO_2$ (FTO) glass, a platinized FTO glass as the cathode and a liquid electrolyte with $I_3^-/I^-$ redox mediator. However, iodine-based electrolyte solution is corrosive, and it absorbs light in the blue part of the visible spectrum [3], while Pt and FTO are expensive. More specifically, the cost of FTO is estimated to be about >20-60% of the cost of the DSC-module [4–7]. Its replacement by cheaper materials should not cause enhancement of the electrode's electrical resistance (typically 10 Ω/sq for FTO) and optical absorption of visible light (typically 20 % near the peak of solar spectrum).

Metallic meshes or metal-wire/polymer-fiber fabric are attractive alternatives for counterelectrodes, offering, moreover, the flexibility as an added practical benefit. An example is platinized-tungsten/PEN fabric (PEN = poly(ethylene 2,6-naphthalate)), which was applied in the 8.5%-efficient I-mediated DSC [8]. Further cost reduction is envisaged upon using cheaper (Pt-free) cathode catalyst and common metals, like stainless steel for the counterelectrode. The replacement of FTO by stainless steel meshes has been attempted also for the $TiO_2$ photoanode [4,5,9–12]. The use of platinized stainless steel for DSC cathodes was reviewed by Toivola et al. [13]. Interestingly, the thermal decomposition of $H_2PtCl_6$ (which is the standard fabrication protocol for Pt@FTO cathode) provided less good catalyst on stainless-steel (SUS304), but sputtered Pt was satisfactorily active on this steel substrate. As many practical DSCs are developed for low illumination intensity (e.g. for indoor use), the high efficiency must be demonstrated at weak light (<0.3 sun) [1,2].

In this work, we addressed all these challenges, and developed a Pt-free and FTO-free counterelectrode for the Co-mediated DSC. The alteration of the $I_3^-/I^-$ redox mediator by Co-based redox couples, like $Co(bpy)_3^{3+}/Co(bpy)_3^{2+}$ provides larger open-circuit voltage of DSC [14–21] and better transparency to visible light [3]. Actually, the state-of-art device with 13% efficiency used this particular mediator in acetonitrile solution and FTO-supported graphene nanoplatelets as the cathode catalyst [14]. Recently, the efficiency was boosted over 14% in a DSC device, using FTO-gold supported graphene nanoplatelets cathode and acetonitrile solution of $Co(phen)_3^{3+/2+}$ redox mediator [21].

Avoiding Pt as a catalyst in Co-mediated DSCs is feasible, because carbons and particularly graphene-based materials provide good electrocatalytic activity for the $Co^{3+}/Co^{2+}$ redox couple [14–21] (for reviews on application of carbons in DSCs see refs. [16,20,22–27]) Although the electrocatalytic activity of optimized carbons (graphenes) is excellent for the $Co^{3+}/Co^{2+}$ reaction (outperforming sometimes the activity of platinum [18]) there is a practical issue that the interfacial adhesion of carbonaceous catalyst to the substrate is usually poor, which presents one of the main limitations of using Pt-free materials in DSC cathodes [25]. A way out seems to be the use of FTO-supported graphene oxide or composites containing graphene oxide [17]. However, this synthetic strategy requires (besides the undesirable FTO) also the catalyst annealing in Ar at high temperatures (450 °C) which are not compatible with fabric substrates containing polymers, like PEN. In this work we have addressed also this problem, and developed a low-temperature (≤200 °C) fabrication protocol towards well-adhering graphene catalysts to stainless steel sheets or wires in a woven fabric. To the best of our knowledge, this kind of cathode and DSC devices are studied here for the first time.

Acetonitrile allows for relatively fast mass transport of the Co-mediator ions, and it was, therefore, used as a solvent in almost all highly-efficient Co-mediated DSCs reported so far [14–20]. Nevertheless, acetonitrile is hardly the best solvent for practical devices, because of its low boiling temperature. There were sporadic trials to replace acetonitrile by less volatile solvents like propionitrile [28], but overcoming the ionic diffusion limitations is still a challenge in Co-mediated DSCs. Also this problem is addressed in this work by a systematic study of less-volatile electrolyte solutions containing sulfolane or ionic liquid in propionitrile. In compliance with practical requirements for e.g. indoor use (where DSCs are competitive to traditional solid-state photovoltaics) [1,2], our study is focused to low-illumination intensities, ≤0.25 sun. We will show that all the above mentioned requirements can be met, i.e. the low-temperature fabrication of mechanically stable, flexible, Pt-/FTO-free cathode for DSC operating with the less-volatile Co-based electrolyte solution is feasible.

## 2. Experimental Section

### 2.1. Materials, Electrodes and Preparation of Electrochemical Cells

Single-layer graphene oxide (GO) and graphene nanoplatelets (GNP), Grade 3 were purchased from Cheap Tubes, Inc. (USA). Stacked graphene platelet nanofiber, acid washed (SGNF) was from ABCR/Strem (Germany). The woven fabric (B23) was from Sefar AG, Switzerland. It consists of transparent PEN fibers (diameter 40 μm) in warp and stainless steel AINSI-316 wires (diameter 31 μm) in weft. Spacing between wires is about 240 μm. The stainless-steel sheet (AISI 316) was from Goodfellow. It was cleaned ultrasonically in isopropanol and acetone. Y123 dye, $Co(bpy)_3(TFSI)_3$ and $Co(bpy)_3(TFSI)_2$ (TFSI = bis((trifluoromethyl)sulfonyl)-imide) were purchased from Dyenamo AB, Sweden. Other chemical were from Aldrich or Merck in the high purity available, and used as received from the supplier. LiTFSI was handled in a glove box under Ar.

GO was dispersed in water by repeated sonication/centrifugation to a concentration of 1 mg/mL. The other graphene-based component (GNO or SGNF) was dispersed in water by sonication and the dispersion was left overnight to separate larger particles by sedimentation. The supernatant containing about 1 mg/mL was stable for several days without marked sedimentation. Precursor for composite electrodes was prepared by mixing this dispersion (GNP or SGNF) with GO solution to a desired proportion of both components.

FTO glass (TEC 15 from Libbey-Owens-Ford, 15 Ohm/sq) was cleaned ultrasonically in isopropanol followed by treatment in UVO-Cleaner (model 256-220, Jelight Co., Inc.). Platinized FTO (Pt@FTO) was prepared by deposition of 5 µL/cm$^2$ of 10 mM $H_2PtCl_6$ in 2-propanol and calcination at 400$^o$C for 15 minutes. The graphene films on FTO were deposited from aqueous solutions by air-brush spraying over a warm substrate (ca. 100$^o$C). The amount of deposited carbon was adjusted by the time of spraying, and was quantified by measurement of optical density. Consistent with our previous works [18,19,29], the optical transmittance at a wavelength of 550 nm, $T_{550}$ served as a parameter characterizing the films. Some electrodes were subsequently annealed in air or Ar at elevated temperatures. The final graphene-based films on FTO had the $T_{550}$ values between 62 and 83 %. The mechanical stability of FTO-supported or steel-sheet-supported catalyst was tested by scratching with a tissue or peeling off by adhesive tape (Scotch). The Sefar B23 fabric was platinized by electrochemical deposition from 10 mM $H_2PtCl_6$ at 2.3 V for 45 s. The average current during electrodeposition was 4 mA/cm$^2$ (normalized to the total grid area). Graphene-based catalyst was deposited on the Sefar B23 grid by dip coating, dried in air and heat-treated for 2 hours in air at elevated temperatures.

The symmetrical sandwich dummy cell was fabricated from two identical FTO sheets which were separated by Surlyn (DuPont; 30-75 µm in thickness) tape as a seal and spacer. The sheet edges of FTO were coated by ultrasonic soldering (Cerasolzer alloy 246, MBR Electronics GmbH) to improve electrical contacts. The exact distance between electrodes was measured by a digital micrometer. The cell was filled with an electrolyte through a hole in one FTO support which was finally closed by a glass/Surlyn seal. Alternatively, the dummy cells were fabricated from two identical stainless steel sheets instead of FTO. The parent electrolyte solution was 0.1 M $Co(bpy)_3(TFSI)_3$ + 0.3 M $Co(bpy)_3(TFSI)_2$ + 0.1 M LiTFSI + 0.2 M tetr-butylpyridine (TBP) in propionitrile. This solution was subsequently mixed with sulfolane or with ionic liquid, 1-ethyl-3-methylimidazolium tetracyanoborate (EMI-TCB) in proportions which are specified in the text.

The DSC with FTO-supported counter electrode was assembled using a Surlyn tape (25 µm in thickness) as a seal and spacer. The DSC with Sefar-grid supported counter electrode was assembled with the 3M End Seal (75 μm thickness) and Surlyn foils, and finally sealed by over-coating with sheet of glass. The edge contacts of grid-based electrodes were improved by silver paint.

Photoelectrochemical tests were carried out with $TiO_2$ films composed of 20 nm diameter anatase. The mesoporous $TiO_2$ films were deposited by screen printing onto FTO (Solar 4 mm thickness, 10 ohms per square, Nippon Sheet Glass). The mesoporous film was 4 μm thick, and was subsequently overcoated with a ~4-μm scattering layer (400 nm diameter, Catalysts & Chemicals Ind. Co. Ltd. (CCIC), HPW-400) and finally post-treated in 0.2 M $TiCl_4$ aqueous solution. The $TiO_2$ electrode was sensitized with Y123 by overnight dipping into 0.1 mM solution of the dye in 4-*tert*-butanol/acetonitrile mixture (1:1 v/v).

## 2.2. Methods

Scanning electron microscopy (SEM) images were obtained by a Hitachi FE SEM S-4800 microscope. The layer thickness was measured by profilometry (Dektak 150, Veeco). Atomic force microscopy (AFM) images were acquired in contact mode using Nanoscope IIIa (Multimode, Bruker) with Si-nitride cantilever NP20 (spring constant nominal $k = 0.16$ N/m). The optical spectra were measured by Perkin Elmer Lambda 1050 spectrometer with integrating sphere in transmission mode. The reference spectrum was air. Electrochemical measurements were carried out using Autolab PGstat-30 equipped with a FRA module (Ecochemie) controlled by the GPES-4 software. Electrochemical impedance data were processed using Zplot/Zview software. The impedance spectra were acquired in the frequency range from 100 kHz to 0.1 Hz, at 0 V bias voltage, the modulation amplitude was 10 mV. For photoelectrochemical tests, the light source was a 450 W xenon light source (Osram XBO 450, Germany) with a filter (Schott 113). The light power was regulated to the AM 1.5G solar standard by using a reference Si photodiode equipped with a color-matched filter (KG-3, Schott) to reduce the mismatch between the simulated light and AM 1.5G to less than 4% in the wavelength region of 350–750 nm. The differing intensities were regulated with neutral wire mesh attenuator. The applied potential and cell current were measured using a Keithley model 2400 digital source meter. The cell active area for illumination was defined by a light-shading mask.

## 3. Results and Discussion

### 3.1. Optimization of Electrolyte Solution

The diffusion of $Co(bpy)_3^{3+}$ (which is the low-concentration component of Co-based electrolyte solutions) is the key process controlling the ionic mass transport in these DSCs. It can be tested using symmetrical dummy cells, which are fabricated from two identical Pt@FTO electrodes with a thin layer of electrolyte solution sandwiched between them [16,30,31]. Cyclic voltammogram (CV) of the reaction:

$$\text{Co(bpy)}_3^{3+} + e^- \rightarrow \text{Co(bpy)}_3^{2+} \quad (1)$$

in this dummy cell exhibits a plateau limiting current density, $j_L$ at sufficient overvoltages, when the current across the cell is diffusion-controlled:

$$j_L = 2nFcD/\delta \quad (2)$$

($n = 1$ is the number of electrons in reaction (1), $F$ is the Faraday constant, $c$ is the concentration of diffusion-limited species, $\text{Co(bpy)}_3^{3+}$, $D$ is the diffusion coefficient and $\delta$ is the distance between electrodes in a dummy cell).

An alternative method for the diffusion-rate assessment is electrochemical impedance spectroscopy (EIS). Depending on the nature of catalyst (see below) the impedance spectra of dummy cells are fitted to two different equivalent circuits (Figure S1 Supporting Info) [16] with $R_S$ = ohmic serial resistance, $R_{CT}$, $R_1$, $R_2$ = charge-transfer resistances, $Z_W$ = Warburg impedance and CPE, $CPE_1$, $CPE_2$ = constant phase elements describing deviation from the ideal capacitance, due to the electrode roughness. In these model circuits, the Warburg impedance equals:

$$Z_W = \frac{W}{\sqrt{i\omega}} \tanh \sqrt{\frac{i\omega}{K_N}} \quad (3)$$

where $W$ is the Warburg parameter and $K_N = (D/0.25\delta^2)$ [16,30,31]. The impedance of constant phase element equals:

$$Z_{CPE} = B(i\omega)^{-\beta} \quad (4)$$

where $B$, $\beta$ are frequency-independent parameters of the CPE ($0 \leq \beta \leq 1$).

Figure S2 (Supporting Info) presents example of electrochemical data for a reference symmetrical dummy cell with two Pt@FTO electrodes. Screening of various electrolyte solutions in dummy cells containing either sulfolane or 1-ethyl 3-methyl imidazolium tetracyanoborate (EMI TCB) ionic liquid is summarized in Figure 1. The reason for addition of sulfolane or EMI TCB to propionitrile solutions is to decrease the electrolyte volatility, which is requested for practical DSC devices. Figure 1 (left chart) shows the limiting current densities obtained from cyclic voltammetry (Eq. 2). The limiting currents are plotted for inter-electrode distance of 30 µm. (Actual experimental data were acquired for variable spacing, $\delta = 30 \pm 3$ µm, but were recalculated to the fixed value of $\delta = 30$ µm for easy comparison). The right chart of Figure 1 presents the diffusion coefficients, found independently from CV (Eq. 2) and EIS (Eq. 3) for all the electrolyte solutions tested. In accord with earlier reports [8,16–19,32] there is reasonable matching of the $D$ values from CV and EIS.

Our maximal diffusion coefficient for pure (additives-free) propionitrile solution is $1.5 \cdot 10^{-6}$ cm$^2$/s, i.e. the ionic transport of Co(bpy)$_3^{3+}$ is about 3-times slower compared to that in the optimized acetonitrile electrolyte solution, which was used previously for high-power (1 sun), 9.3%-efficient solar cells ($D = 4.6 \cdot 10^{-6}$ cm$^2$/s) [18]. Addition of sulfolane or EMI TCB causes a considerable drop of the mass-transport rate (Fig. 1). Nevertheless, these low-volatile, viscous electrolyte solutions are still applicable for DSCs operating at low light intensities. For instance, at 0.1 sun, we expect photocurrents of *ca.* 2 mA/cm$^2$ in good cells [2], and this current is supported by ionic-transport even with the 50 vol% additive in propionitrile (Fig. 1). Furthermore, larger currents can be achieved by decreasing of the inter-electrode spacing ($\delta$) in DSC (*vide infra*). For subsequent tests of catalysts and the DSCs optimization, reported below, we have chosen the electrolyte solution with 11 vol % sulfolane as the reference medium.

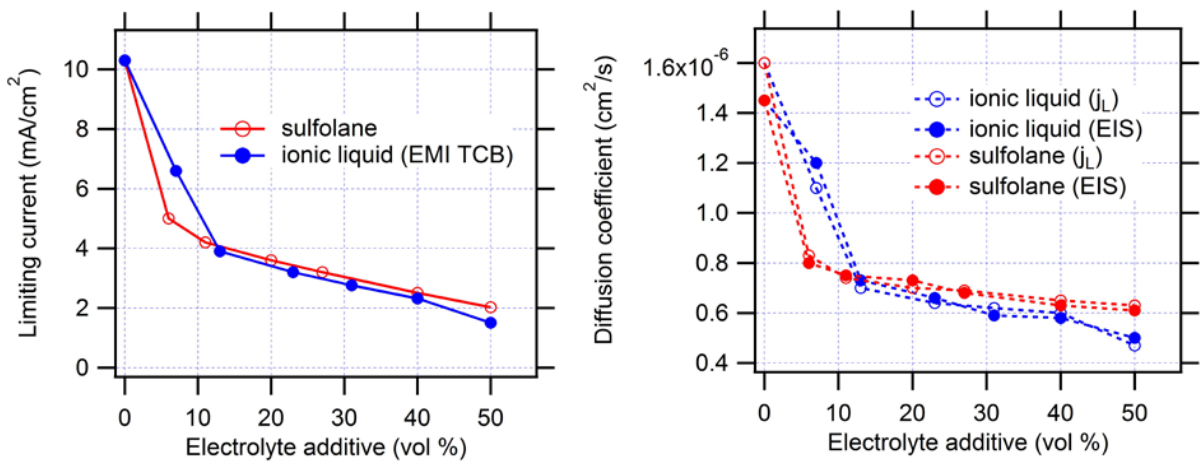

**Figure 1.** Limiting current of Co(bpy)$_3^{3+}$ transport (for δ = 30 µm) from cyclic voltammetry and the corresponding diffusion coefficient in electrolyte solutions composed of propionitrile solvent with varying amount of sulfolane (red curves) or ionic liquid (EMI TCB; blue curves). Diffusion coefficients were obtained either from cyclic voltammetry (j$_L$) or from impedance spectroscopy (EIS).

### 3.2. Optimization of cathode catalyst

Co(bpy)$_3^{3+}$ is produced in a DSC through dye regeneration near the illuminated anode, and is subsequently transported to the cathode by diffusion. Here, it is reduced back at a rate, which is ideally equal or higher than the rate of the Co(bpy)$_3^{3+}$ generation and its transport across the DSC. This quantifies the necessary exchange current density at the cathode, $j_o$ to be comparable with the photocurrent density on the TiO$_2$ photoanode. The latter is *ca.* 20 mA/cm$^2$ in good cells under 1 sun illumination [2]. The exchange current density equals:

$$j_0 = \frac{RT}{nFR_{CT}} = Fk_0(c_{ox}^{1-\alpha} \cdot c_{red}^{\alpha}) \qquad (5)$$

$R$ is the gas constant, $T$ is temperature, $n$ is the number of electrons in the electrode reaction, $F$ is Faraday constant, $R_{CT}$ is the charge-transfer resistance, $k_0$ is the formal (conditional) rate constant of the electrode reaction, $c_{ox}$ and $c_{red}$ are the concentrations of oxidized and reduced mediator, respectively and α is the charge-transfer coefficient (α ≈ 0.5). For $j_0 = 5$ mA/cm$^2$ (produced under 0.25 sun illumination, see above) we calculate $R_{CT} = 5.2$ Ωcm$^2$ to be the desired value of charge-transfer resistance at these conditions.

Fitting the EIS data in Fig. S2 provides a well-acceptable value of $R_{CT} = 2.5$ Ωcm$^2$ for the standard, thermally made, Pt@FTO cathode in the tested electrolyte solution. Nevertheless, replacing the Pt@FTO by a PEN/steel cathode requires alternative low-temperature fabrication protocols to be developed. One of potential options is the catalysts based on graphene oxide (GO) [17]. Figure 2 surveys our initial screening tests of a pure GO-catalyst. To reproduce the conditions of our earlier study [17], we started our tests with the optimum material, which was GO on FTO, heat-treated in Ar-atmosphere at 450 °C for 1 hr. The corresponding EIS spectrum in Fig. 2 provided $R_{CT} = 2.6$ Ωcm$^2$, which is only slightly worse than the value for Pt@FTO (see above) and the corresponding values from Ref. [17] (0.51-1.39 Ωcm$^2$, which were, however, measured in another electrolyte solution; acetonitrile-based).

More importantly, active catalysts can be prepared by calcination (2 hrs) of GO at lower temperatures in air, with $R_{CT}$ of 3.0, 5.0 and 7.9 Ωcm$^2$ at the calcination temperatures of 300, 250 and 200 °C, respectively, but there is a sudden jump to lower catalytic activity for GO heat-treated below 200 °C (Fig. 2). In all cases, good adhesion of the heat-treated GO to FTO substrate was found. The adhesion was quantified by the $T_{550}$ values which were measured after abrasion by a tissue or after delamination with an adhesive tape (Scotch). Scratching by a tissue caused ≈20% drop of the optical density at 550 nm wavelength, but the Scotch tape test caused about 80% drop of the optical density (see Figure S3 in Supporting Info). Detaching by Scotch tape left on FTO a thin, firmly-adhering residual layer ($GO_{res}$). The edges between pristine GO, $GO_{res}$ and pure FTO (Fig. S4 in Supporting Info) were further investigated by a contact-mode atomic force microscopy (AFM). The results are shown in Figures S5 and S6 (Supporting Info). The AFM profile analysis indicates the thickness of $GO_{res}$ to be about 20 % of the original thickness of the pristine GO film. This proportion is reasonably matching the corresponding ratio of optical densities ($GO_{res}/GO \approx 17$ %) measured by UV-Vis spectra (Fig. S3). The $GO$-$GO_{res}$ edge is not significantly damaged by an AFM tip at pressures as high as ≈ 0.15 GPa (Fig. S6 in Supporting Info).

To the best of our knowledge, the low-temperature activation of GO for the $Co^{3+/2+}$ electrocatalysis is shown here for the first time. Ho et al. [33] recently reported on low-temperature (150-250$^o$C) activation of GO@FTO for DSCs with I-mediator, but the performance of their catalysts was not good compared to that of the control system with Pt@FTO. Obviously, GO is less suitable catalyst for the $I_3^-/I^-$ couples than for the $Co^{3+/2+}$ couples.

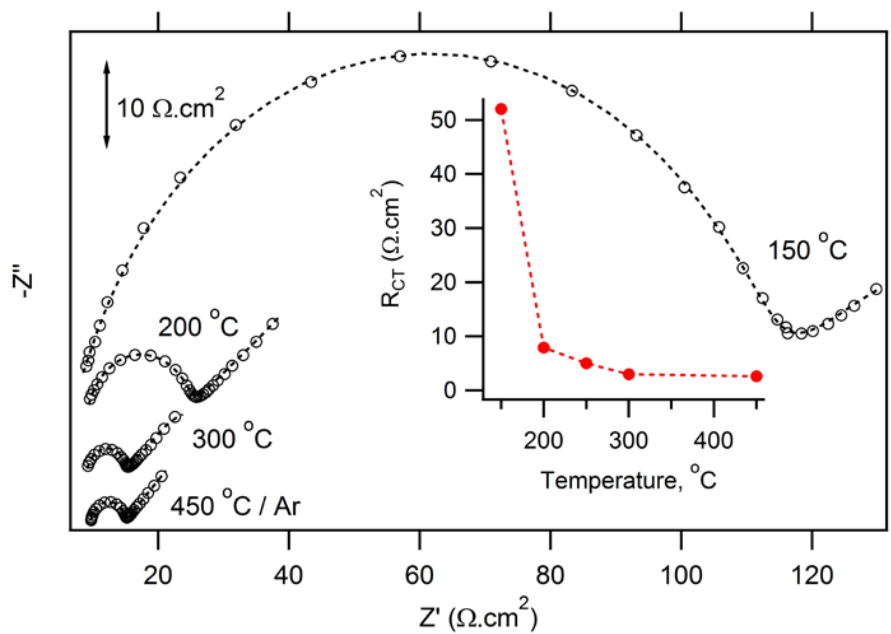

**Figure 2:** Nyquist plot of electrochemical impedance spectra measured at 0 V from 100 kHz to 0.1 Hz on symmetrical dummy cells. The cells were from two identical FTO electrodes (75 um spacing) covered by graphene oxide film heat treated at 150 C$^o$ - 300 $^o$C in air (or in Ar atmosphere at 450 $^o$C). Experimental impedances (points) were fitted to equivalent circuit, model 1, Fig. S1 (dashed lines). Electrolyte solution: 0.1 M Co(bpy)$_3$TFSI$_3$ + 0.3 M Co(bpy)$_3$TFSI$_2$ + 0.1 M LiTFSI + 0.2 M TBP in propionitrile + 11 vol % sulfolane. Inset: Charge transfer resistance ($R_{CT}$) of graphene oxide heat-treated at various temperatures.

Alternative highly-active catalysts are based on graphene nanoplatelets (GNP) [14,18,19] or stacked graphene platelet nanofibers (SGNF) [28]. The corresponding impedance spectra in FTO-dummy cells are shown in Figures 3a and 3b. Both these materials are excellent electrocatalysts, even without thermal activation (the used temperature of 110$^o$C is just for spray deposition and drying). The fitted $R_{CT}$ values are: 0.56 Ωcm$^2$ and 0.32 Ωcm$^2$ for GNP and SGNF, respectively. The last mentioned $R_{CT}$ (for SGNF) is the best charge-transfer resistance observed in this work. Unfortunately, the adhesion of the catalyst (GNP or SGNF) to the substrate is bad, in accord with earlier experience [17,18]. To address this problem, we further tested composite materials containing GO in a mixture with either GNP or SGNF. These composites are coded GONPx or GONFx, respectively (x denotes the concentration of GNP or SGNF in wt%). Representative data are shown in Figure 4.

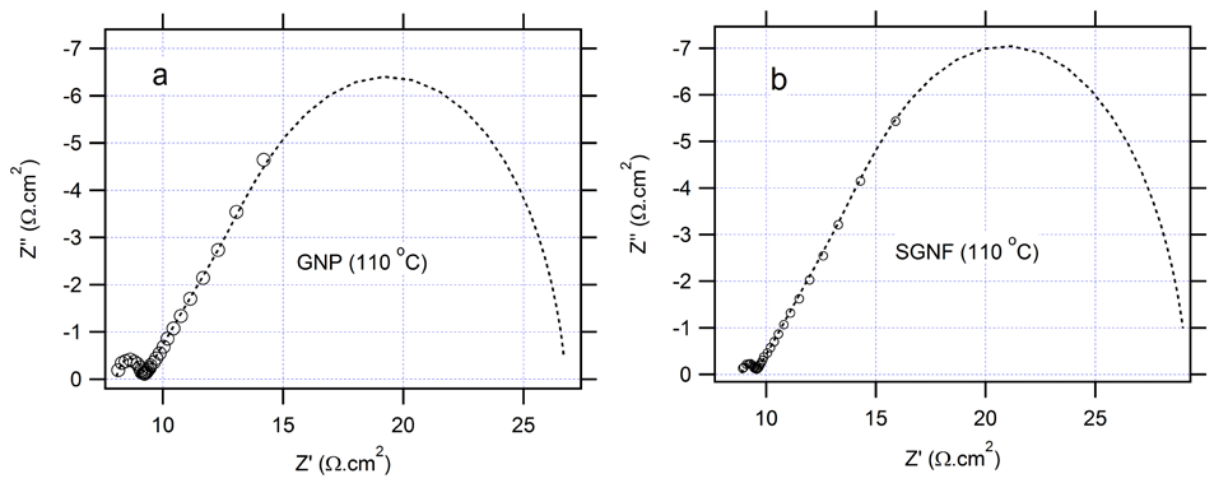

**Figure 3.** Nyquist plot of electrochemical impedance spectra measured at 0 V from 100 kHz to 0.1 Hz on symmetrical dummy cells. The cells were from two identical FTO electrodes (75 um spacing) covered by graphene nanoplatelets (GNP, left chart) or stacked graphene platelet nanofibers (SGNF, right chart) heat treated at 110 °C in air. Experimental impedances (points) were fitted to equivalent circuit, model 1, Fig. S1 (dashed lines). Electrolyte solution: 0.1 M $Co(bpy)_3(TFSI)_3$ + 0.3 M $Co(bpy)_3(TFSI)_2$ + 0.1 M LiTFSI + 0.2 M TBP in propionitrile + 11 vol % sulfolane.

The EIS spectra of composite electrodes need to be fitted to a more complex equivalent circuit (model 2; Fig. S1). It is based on a serial combination of two R-CPE elements, originating from two distinct subunits: one directly interacting with FTO ($R_1$-$CPE_1$) and the other ($R_2$-$CPE_2$) being interfaced in series [17]. The found charge transfer resistances $R_1$&$R_2$ equal (in $\Omega cm^2$): 1.7&5.4, 5.2&5.3, 1.7&4.1, 0.5&1.4 for GONP50, GONF50, GONF70, GONF80, respectively (cf. Fig. 4). All these composite catalysts exhibit good adhesion to the substrate. Obviously, the surface-bonding component is here the graphene oxide, which interacts with FTO through oxygen-containing functional groups present at both surfaces. Furthermore, even at temperatures below 200 °C, the additional carbonaceous component (GNP or SGNF) is firmly cemented in the catalytic layer by GO, presumably again through oxygen-containing defects in the carbonaceous skeleton.

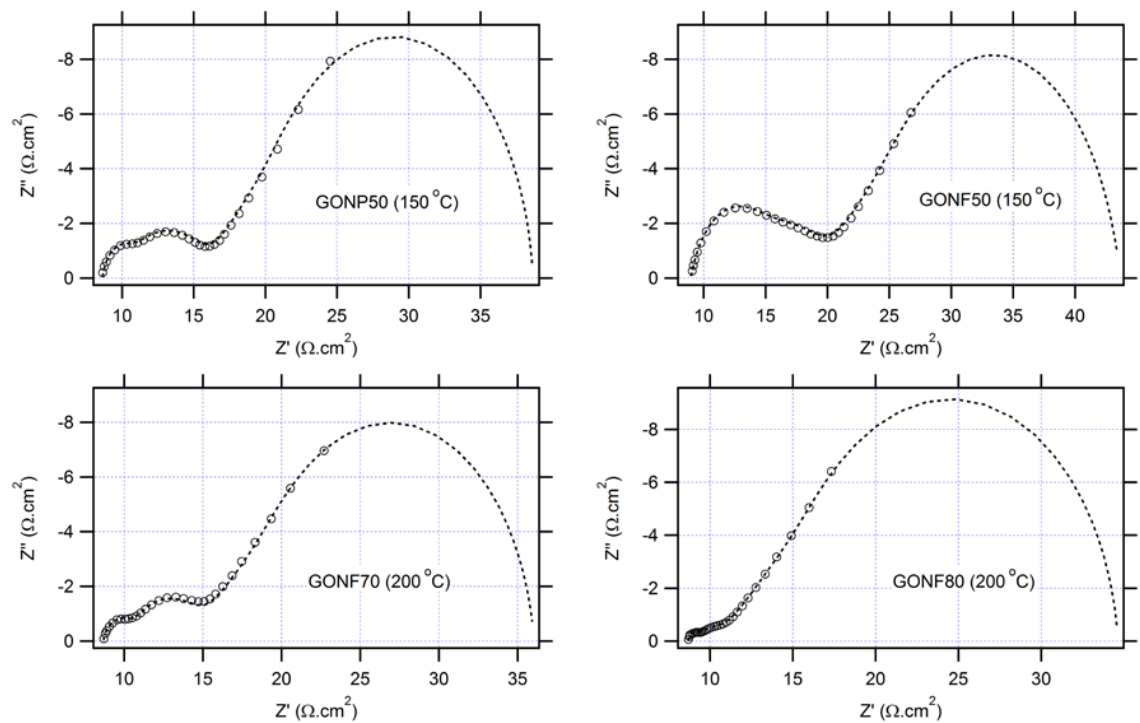

**Figure 4**. Nyquist plot of electrochemical impedance spectra measured at 0 V from 100 kHz to 0.1 Hz on symmetrical dummy cells. The cells were from two identical FTO electrodes (75 um spacing) covered by a composite of graphene oxide and graphene nanoplatelets (GONPx) or stacked graphene platelet nanofibers (GONFx); (x denotes the concentration of GNP or SGNF in wt%). The heat-treatment temperature is labeled in annotations. Experimental impedances (points) were fitted to equivalent circuit, model 2, Fig. S1 (dashed lines). Electrolyte solution: $Co(bpy)_3(TFSI)_3$ + 0.3 M $Co(bpy)_3(TFSI)_2$ + 0.1 M LiTFSI + 0.2 M TBP in propionitrile + 11 vol % sulfolane.

In addition to FTO, also stainless steel is a suitable substrate for our GO-containing catalysts. Figure 5 presents an example plot for symmetrical dummy cell from AINSI 316 stainless steel sheet, which was coated by graphene oxide activated at 200°C. Fitting of this impedance spectrum to model 1 (Fig. S1) equivalent circuit provided $R_{CT}$ = 6.9 $\Omega cm^2$, which is close to the $R_{CT}$ value of the same catalyst on FTO substrate (7.9 $\Omega cm^2$), see Fig. 2 and discussion thereof. (The slightly better value for steel electrode could be just coincidental). Although both FTO and steel were coated by GO at identical deposition conditions (time, temperature, position in the spray-deposition reactor), the catalyst loading on steel sheet cannot be quantified by its optical transmittance, $T_{550}$, like on FTO, see Experimental Section). The steel-supported electrodes show expectedly the best serial resistance ($R_s$ = 0.8 $\Omega$cm$^2$; Fig. 5) whereas the FTO-supported electrodes provided the $R_s$ values between 4.0 and 4.6 $\Omega$cm$^2$ (from Figs 2-4). Similarly to FTO, we note good adhesion of GO and GO-containing composites to the steel surface.

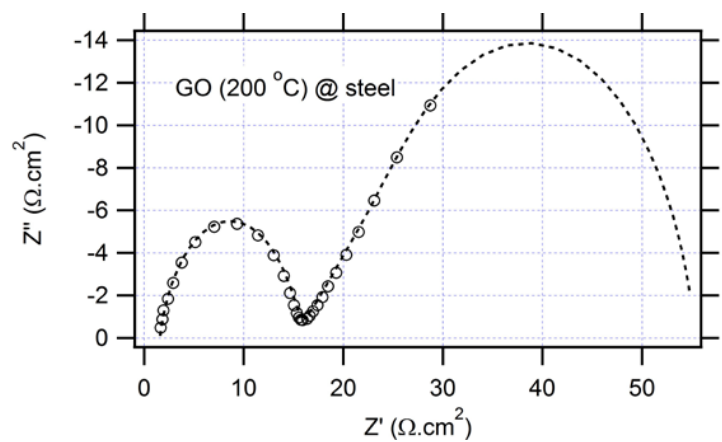

**Figure 5**. Nyquist plot of electrochemical impedance spectra measured at 0 V from 100 kHz to 0.1 Hz on symmetrical dummy cells. The cells were from two identical stainless steel (AINSI 3126) electrodes (75 um spacing) covered by graphene oxide heat-treated at 200°C in air. Experimental impedances (points) were fitted to equivalent circuit, model 1, Fig. S1 (dashed lines). Electrolyte solution: 0.1 M Co(bpy)$_3$(TFSI)$_3$ + 0.3 M Co(bpy)$_3$(TFSI)$_2$ + 0.1 M LiTFSI + 0.2 M TBP in propionitrile + 11 vol % sulfolane.

### 3.3. Dye-sensitized Solar Cells

Reference dye-sensitized solar cells had the traditional architecture of two plan-parallel FTO sheets with a thin layer of electrolyte solution sandwiched between them. The photoanode was TiO$_2$ sensitized with 3[6-[4-[bis(2',4'-dihexyloxybiphenyl-4-yl)amino-]phenyl]-4,4-dihexyl-

cyclopenta-[2,1-b:3,4-b']dithiophene-2-yl]-2-cyanoacrylic acid, coded **Y123** which is one of the commonly used dyes for the Co-mediated DSCs [15,17–19] (see Supporting Info, Scheme S7 for chemical formula of **Y123**). The counter-electrode activity can be tested on a complete DSC, too. Figure 6 presents example data, which were fitted to the equivalent circuit of DSC (Figure S8 in Supporting Info). At the used experimental conditions (forward bias near $V_{oc}$ in dark) we observe distinguished high-frequency semicircle assignable to the counter-electrode. The second semicircle is assigned to recombination resistance at $TiO_2$/electrolyte solution interface, which is parallel to $CPE_\mu$, modelling the chemical capacitance in $TiO_2$ (transport resistance in $TiO_2$ is negligible at these conditions). The third (lowest frequency) arc is assigned to Warburg impedance in the electrolyte solution.

The found counter-electrode's $R_{CT}$ values were 2.6 $\Omega cm^2$ (for Pt@FTO cathode) and 3.2 $\Omega cm^2$ (for GONP50@FTO cathode) from the spectra in Fig. 6. The Pt-cathode in DSC exhibits similar charge-transfer resistance to that determined in the corresponding dummy cell (2.5 $\Omega cm^2$; see Fig. S1 and discussion thereof). Our values are comparable or better than those reported by Jeon et al. [34], who had analogously measured EIS both on their dummy cells and DSCs. In the case of GONP50, the situation is more complex, because fitting of the highest-frequency semicircle in the DSC spectrum to a circuit of composite cathode (cf. model 2 in Fig. S1) was not reliable in this case. Nevertheless, the overall average value of charge-transfer resistance, $R_{CT}$ found for the GONP50-based DSC cathode (3.2 $\Omega cm^2$) is not far from the resistances $R_1$&$R_2$ (1.7&5.4 $\Omega cm^2$) observed for the dummy cell (cf. Fig. 4a). This comparison also demonstrates, that the symmetrical dummy cells allow for more accurate analysis of subtle effects on the counterelectrode than the complete DSC.

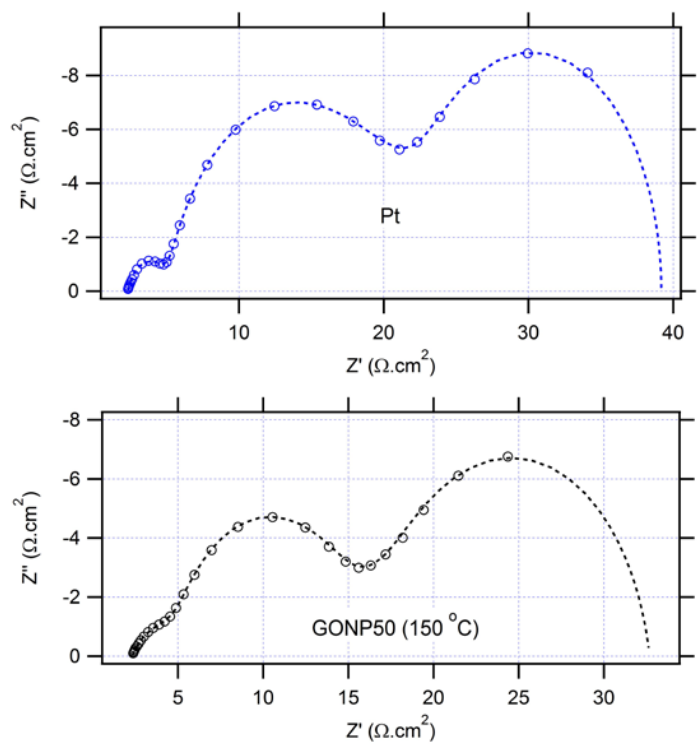

**Figure 6**. Nyquist plot of electrochemical impedance spectra measured at 0 V from 100 kHz to 0.1 Hz on complete DSC (electrode spacing 23-25 µm) with either Pt@FTO cathode (top chart) or GONP50@FTO cathode (bottom chart) heat-treated at 150$^{o}$C in air. Experimental impedances (points) were fitted to equivalent circuit Fig. S8 (dashed lines). Electrolyte solution: 0.1 M Co(bpy)$_3$(TFSI)$_3$ + 0.3 M Co(bpy)$_3$(TFSI)$_2$ + 0.1 M LiTFSI + 0.2 M TBP in propionitrile + 11 vol % sulfolane.

Figure 7 shows the current/voltage characteristics of dye-sensitized solar cell with either Pt@FTO (thermally made) cathode or GONF70@FTO cathode (GONF70 was heat treated at 200 $^{o}$C). The DSCs were illuminated with low-intensity light corresponding to 10 and 25 % sun. Detailed data from solar tests are summarized in Table 1. They confirm that DSC with graphene-based cathode (GONF70) exhibits comparable or even slightly better (at 0.25 sun) efficiency compared to that for DSC with thermally made Pt@FTO cathode. Furthermore, the dark current is almost identical for both cathode types, which is an indirect proof of firm adhesion of the catalyst to cathode substrate [17, 18].

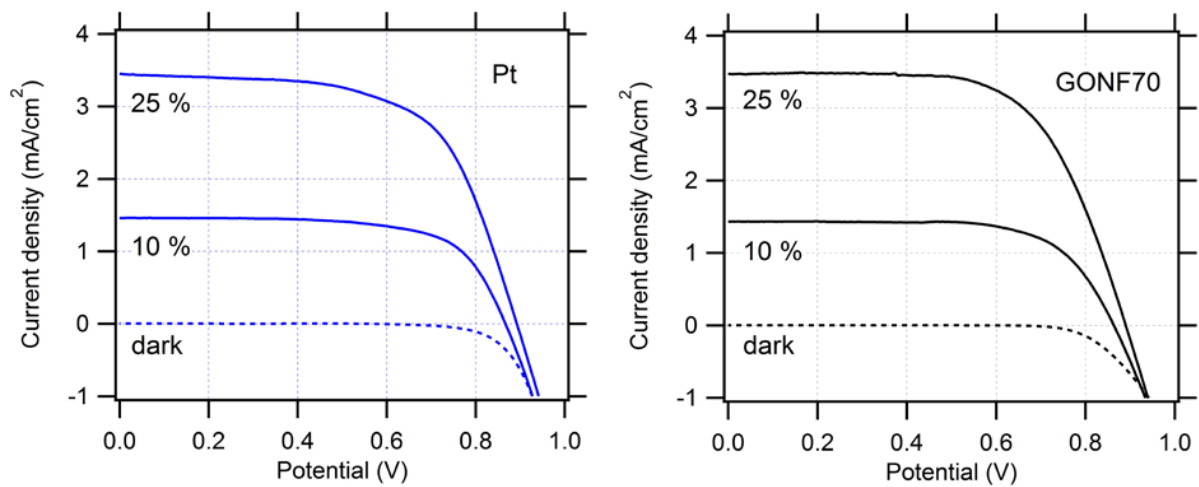

**Figure 7.** Current-voltage characteristics of dye sensitized solar cells with Y-123 sensitized $TiO_2$ photoanode. Electrolyte solution: 0.1 M $Co(bpy)_3(TFSI)_3$ + 0.3 M $Co(bpy)_3(TFSI)_2$ + 0.1 M LiTFSI + 0.2 M TBP in propionitrile + 11 vol % sulfolane. Left chart is for DSC with Pt-cathode; right chart is for DSC with GONF70 cathode, heat-treated at 200°C ($T_{550}$ = 78%). The illumination intensity is 25% sun, 10% sun and 0 (curves dark).

Finally, we tested the performance of our steel/PEN-fabric counterelectrodes in dye-sensitized solar cell. There are three main advantages of the fabric-based electrodes (over FTO): cheaper price, better conductivity and better optical transmittance [8]. The structure of our Sefar-B23 fabric is shown on Figure S9 (Supporting info). It depicts the SEM images of the fabric, which was coated by graphene oxide or GONF80 and activated at 200°C in air. Obviously GO is deposited in irregular islands on the PEN fibers, which are visualized through differential charging during SEM measurement. On the other hand, the coating of stainless-steel wire by GO is perfectly homogeneous, as shown on the detailed view of the wire surface (Figure S9) and also on its energy-dispersive X-ray (EDX) spectroscopic mapping (Figure S10). Contrarily, the GONF80-coated wire exhibits distinct morphology with fibrous particles well adhering to the surface (Figures S9 and S10).

Our grids had low sheet resistance (< 1 Ω/sq in the direction of the steel wires) which is significantly outperforming the resistivity of FTO (10-15 Ω/sq). Optical transmittance is also better for the naked pure grids, although the carbon coating causes its decrease by about 10 %, see Figure S11 in Supporting Info. (The coating of PEN causes unwanted darkening of the fabric, too, but it is unavoidable in the dip-coating deposition). Another concern associated with a grid electrode is the complicated mass-transport geometry [8].

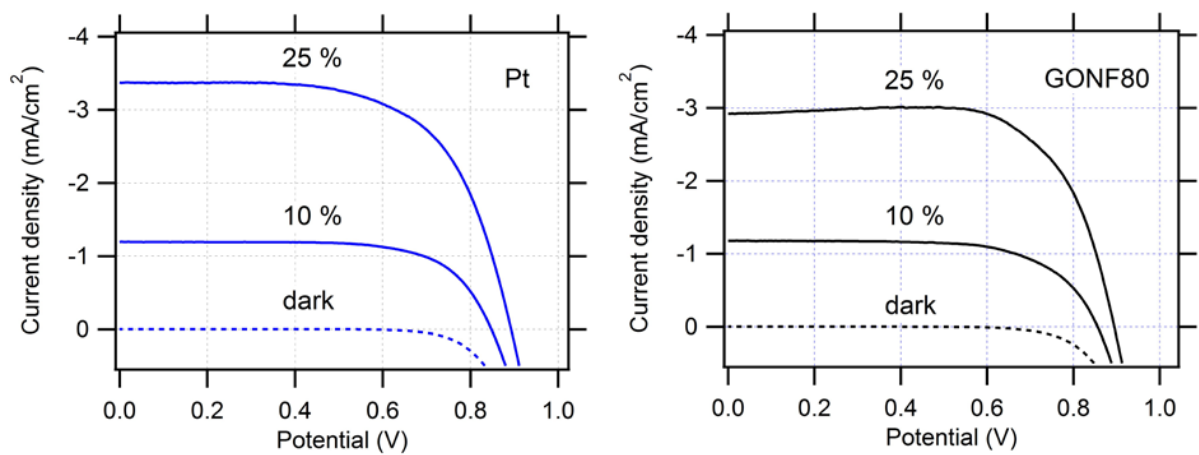

Figure 8. Current-voltage characteristics of dye sensitized solar cell with sensitized $TiO_2$ photoanode and stainless steel/PEN-grid cathode (B23). Electrolyte solution: 0.1 M $Co(bpy)_3(TFSI)_3$ + 0.3 M $Co(bpy)_3(TFSI)_2$ + 0.1 M LiTFSI + 0.2 M TBP in propionitrile + 11 vol % sulfolane. Left chart is for DSC with Pt-cathode (Pt deposited electrochemically); right chart is for DSC with GONF80 cathode, heat-treated at 200°C. The illumination intensity is 25% sun, 10% sun and 0 (curves dark).

Figure 8 shows the corresponding current/voltage characteristics of DSCs with steel/PEN-grid-based cathode, either platinized of made from graphene-based composite. To avoid thermal degradation of PEN, the catalyst loading was carried out at low-temperature by (1) electrochemical deposition of Pt [8] or (2) GONP80 activation at 200$^{o}$C, respectively. Detailed data from DSCs testing are compiled in Table 1. It turns out that both Pt and graphene-based catalysts behave similarly in the actual DSCs, independent of the used cathode support (FTO or steel wire in a fabric). The overall slightly smaller efficiency of fabric-based DSCs is presumably caused by hindered mass transport in the fabrics. Again, we note negligible differences in the dark current for Pt- and GONF80-catalysts on fabric supported cathodes, which is indirect argument in favor of good catalyst adhesion [17, 18].

**Table 1.** Characteristics of solar cells with Y123-sensitized $TiO_2$ photoanodes at 0.1 sun and 0.25 sun illumination. Electrolyte solution: 0.1 M $Co(bpy)_3(TFSI)_3$ + 0.3 M $Co(bpy)_3(TFSI)_2$ + 0.1 M LiTFSI + 0.2 M TBP in propionitrile + 11 vol % sulfolane. Short circuit photocurrent density = $j_{SC}$, open-circuit voltage = $V_{OC}$, fill factor = $FF$, solar conversion efficiency = $\eta$. The cell active area was 0.3 $cm^2$.

| Counterelectrode | Illumination (sun) | $j_{SC}$ (mA/cm$^2$) | $V_{OC}$ (mV) | FF | $\eta$ (%) |
|---|---|---|---|---|---|
| Pt@FTO (thermal) | 0.1 | 1.46 | 867 | 0.68 | 8.6 |
| Pt@FTO (thermal) | 0.25 | 3.45 | 893 | 0.63 | 7.7 |
| GONF70@FTO (200°C) | 0.1 | 1.43 | 864 | 0.68 | 8.5 |
| GONF70@FTO (200°C) | 0.25 | 3.47 | 890 | 0.64 | 7.9 |
| Pt@B23 (electrodep.) | 0.1 | 1.20 | 848 | 0.69 | 7.5 |
| Pt@B23 (electrodep.) | 0.25 | 3.37 | 893 | 0.64 | 7.1 |
| GONF80@B23 (200°C) | 0.1 | 1.18 | 856 | 0.66 | 7.0 |
| GONF80@B23 (200°C) | 0.25 | 2.92 | 895 | 0.69 | 6.9 |

## 4. Conclusions

Propionitrile mixed with sulfolane or with EMI TCB ionic liquid is an acceptable less-volatile solvent for formulating the electrolyte solutions for $Co(bpy)^{3+}/Co(bpy)^{2+}$-mediated DSCs. Although the diffusion coefficient of $Co(bpy)_3^{3+}$ is ca. 3-times smaller in propionitrile solutions (referenced to acetonitrile solutions) it is still sufficient to support the adequately fast ionic charge-transport in DSCs working at low illumination intensity (≤0.25 sun).

Highly-active catalysts for the electrochemical charge-transfer in the $Co(bpy)^{3+}/Co(bpy)^{2+}$ couple can be prepared at temperatures as low as ≤200°C. The catalytic layers are well adhering to FTO or steel surfaces, both flat steel sheet and steel wires in the woven fabric. The woven fabric consisting of transparent polyester (PEN) fibers in warp and stainless steel wires in weft is applicable as a flexible cathode in liquid-junction dye-sensitized solar with the $Co(bpy)^{3+}/Co(bpy)^{2+}$ redox mediator.

The dye-sensitized solar cells with various cathodes, fabricated either from Pt or from optimized graphene-based catalysts, and supported by either FTO or by stainless-steel/PEN fabric, respectively show similar performance. The solar conversion efficiencies are between 6.9 and 7.9 % in all four cathode variants at 0.25 sun illumination. This confirms that the stainless-steel/PEN woven fabric is promising replacement of conductive glass in the DSC counterelectrodes.

**Acknowledgements**

This work was supported by the Swiss Commission for Technology and Innovation (CTI) project No. 16452.2 PFNM-NM and by the European Research Council through the Advanced Research Grant no. 247404 'Mesolight' and by the European Union FP7 Programme (No. 604391 Graphene Flagship). L.K. acknowledges the support from Grant Agency of the Czech Republic (contract No. 13-07724S). Thanks are due to Dr. Peter Chabrecek (Sefar AG) for providing the grid samples, to Jean-David Décoppet (LPI-EPFL) for technical assistance with solar tests, to Pavel Janda (JHIPC) for AFM measurements and to Milan Bousa (JHIPC) for SEM/EDX measurements.
**Supporting Information**

Equivalent circuit for EIS fitting, additional electrochemical data on dummy cells, formula of Y123, optical spectra, AFM images, SEM/EDX data on B23-fabric.